\UseRawInputEncoding
\documentclass[aps,pra,reprint,amsmath, amssymb,superscriptaddress]{revtex4-2}
\usepackage{xcolor}
\usepackage{graphicx}
\usepackage{dcolumn}
\usepackage{bm}
\usepackage{physics}
\usepackage{braket}
\usepackage{soul}
\usepackage{ulem}
\begin{document}

\preprint{APS/123-QED}

\title{Sympathetic Cooling of Levitated Optomechanics through Nonreciprocal Coupling}
\author{Jialin Li}
\author{Guangyu Zhang}
\author{Zhang-qi Yin}
    \email{zqyin@bit.edu.cn}
    \affiliation{Center for Quantum Technology Research and Key Laboratory of Advanced Optoelectronic Quantum Architecture and Measurements (MOE), \\ School of Physics, Beijing Institute of Technology, Beijing 100081, China}

\date{\today}

\begin{abstract}

Optomechanical cooling of levitated nanoparticles has become an essential topic in modern quantum physics, providing a platform for exploring macroscopic quantum phenomena and high-precision sensing. However, conventional cavity-assisted cooling is fundamentally constrained by cavity dissipation and environmental noise, limiting the attainable minimum temperature. In this work, we propose a non-Hermitian optomechanical cooling scheme through nonreciprocal coupling between two levitated nanoparticles, where one particle is directly cooled by an optical cavity and the other is cooled indirectly through a non-Hermitian interaction. Both analytical solutions and numerical simulations reveal that increasing nonreciprocity enhances directional energy transfer, enabling the target particle to reach a lower phonon occupation than is achievable in conventional cavity cooling. This study demonstrates a new cooling mechanism driven by non-Hermitian interactions, offering theoretical guidance for realizing controllable energy flow and deep cooling in levitated optomechanical systems, and paving the way for future developments in quantum control and sensing technologies.

\end{abstract}

\maketitle


\section{\label{sec:level1}Introduction}

Optomechanics, which explores the interaction between optical fields and mechanical motion, has emerged as a pivotal platform for realizing fundamental quantum control and precision measurement at the mesoscopic scale~\cite{Aspelmeyer2014,Kippenberg2008}. It plays a central role in diverse applications such as optical cooling, quantum-state manipulation, and high-sensitivity force detection. In recent years, optical trapping and manipulation of levitated nanoparticles have attracted increasing attention due to their exceptional isolation from the environment and potential for observing macroscopic quantum phenomena, nonlinear dynamics, and ultrasensitive metrology~\cite{Rudolph2020,Millen2020,Stickler2020,Barker2010}. Among various techniques, cooling levitated particles by coupling them to an optical cavity has become a cornerstone method that enables the suppression of thermal motion through cavity-mediated energy dissipation~\cite{Gonzalez-Ballestero2019,Delić2020}. Such schemes have been successfully implemented in a variety of systems, including levitated microspheres, nanoparticle arrays, and atomic ensembles, where the coupling between cavity and mechanical modes allows energy transfer from mechanical degrees of freedom to the optical field, effectively reducing the temperature of the particle mode~\cite{Gieseler2012,Anetsberger2009}.

Despite these advances, the achievable cooling limit in conventional cavity optomechanics remains restricted by cavity losses, coupling strength, and the intrinsic properties of the mechanical element~\cite{WilsonRae2007,Genes2008}. In levitated systems, especially under ultrahigh vacuum conditions, the final phonon occupation is typically constrained by the balance between cavity dissipation and environmental noise~\cite{Delić2020}. These fundamental limitations have motivated the search for new cooling mechanisms that can surpass the traditional sideband or feedback cooling limits, including unconventional dissipative engineering and multi-mode strategies~\cite{Mari2012,Ma2016}.

Recently, the introduction of non-Hermitian physics into photonic and optomechanical systems has opened new avenues for controlling energy flow and dissipation. Non-Hermitian interactions, characterized by asymmetric or nonreciprocal coupling, can induce directional energy transport and an effective gain–loss balance within composite systems. Such interactions have enabled a range of intriguing phenomena, including unidirectional photon(phonon) transmission~\cite{Regensburger2012,Chen2017,Liu2020}, exceptional points~\cite{Feng2017,Doppler2016}, and topological mode control~\cite{ElGanainy2018,Liu2020}. However, their potential for manipulating thermalization processes and achieving deep cooling in mechanical systems has not yet been thoroughly explored. Recent theoretical proposals suggest exponentially enhanced cooling using non-Hermitian chains~\cite{Xu2024} and related reservoir-engineering approaches~\cite{Metelmann2015,Ruesink2016}.

In this work, we propose a scheme for sympathetic cooling through nonreciprocal coupling between two optically levitated nanoparticles. In our model, only one particle (the auxiliary particle) is directly coupled to an optical cavity, while the other particle (the target particle) interacts with the auxiliary particle through a nonreciprocal mechanical coupling. The cavity–particle coupling enables standard optomechanical cooling for the auxiliary particle, while the non-Hermitian interparticle interaction allows directional energy transfer from the target particle to the auxiliary particle, effectively extending the cooling effect to the target particle. By tuning the nonreciprocal interaction strength, we demonstrate that the target particle can reach a lower steady--state phonon number than would be achievable through direct cavity cooling alone.

The remainder of this paper is organized as follows. In Sec.~\uppercase\expandafter{\romannumeral2}, we present the model and the non-Hermitian Hamiltonian of the coupled system. In Sec.~\uppercase\expandafter{\romannumeral3}, we derive the analytical solution by first obtaining the cavity-induced effective damping and temperature of the auxiliary particle, and then incorporating them into a two-mode non-Hermitian model of the two particles. In Sec.~\uppercase\expandafter{\romannumeral4}, we perform numerical simulations based on the full master equation to investigate the dynamical cooling behavior and verify the analytical results. Finally, Sec.~\uppercase\expandafter{\romannumeral5} concludes with a discussion of the implications of our findings and potential applications of non-Hermitian cooling in levitated optomechanics and quantum thermodynamic control.

\section{\label{sec:level2}The Scheme}

We consider two optically levitated nanoparticles coupled through a nonreciprocal interaction and an optical cavity coupled only to one of them, as shown schematically in Fig.~1. Each nanoparticle is confined by a linearly polarized Gaussian trapping laser and behaves as a quantum oscillator with frequencies $\omega_a$ and $\omega_b$ for particles $A$ and $B$, respectively. An optical cavity mode with resonance frequency $\omega_c$ interacts with particle~$A$ through radiation pressure coupling. particle~$B$, in contrast, does not couple directly to the cavity field but interacts with particle~$A$ through a nonreciprocal coupling channel, which can be engineered by controlling the optical phase, laser wavelength, cavity geometry, and the spatial separation between the particles.

\begin{figure}[htbp]
\includegraphics[width=0.5\textwidth]{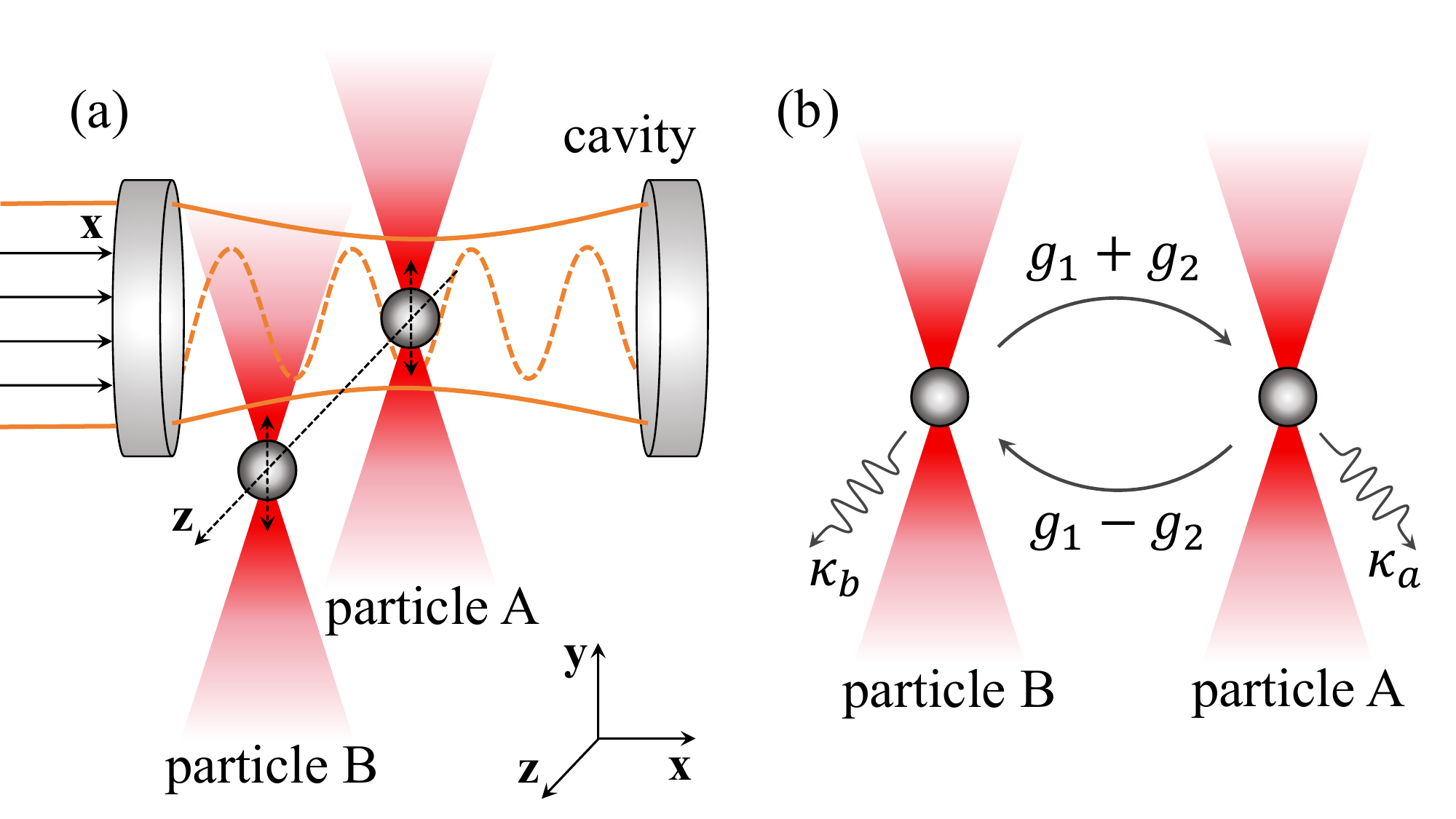}
\caption{\label{fig:fig1} (a) Schematic diagram of the nonreciprocal optomechanical cooling system. particle~$A$ is optically trapped inside the cavity and coupled to the cavity mode through radiation pressure, while particle~$B$ is levitated outside the cavity. Each particle supports a quantized harmonic mode with frequencies $\omega_a$ and $\omega_b$, respectively. (b) Illustration of the nonreciprocal coupling between particles $A$ and $B$ and their mechanical dissipation processes. The coupling strengths and mechanical decay rates are explicitly labeled.}
\end{figure}

The total Hamiltonian of the system can be expressed as $H = H_0 + H_1$. The non-interaction part is given by $H_0 = \omega_c c^{\dagger}c + \omega_a a^{\dagger}a + \omega_b b^{\dagger}b$, where $a\ (a^{\dagger})$, $b\ (b^{\dagger})$, and $c\ (c^{\dagger})$ denote the annihilation (creation) operators of particle~$A$, particle~$B$, and the cavity mode, respectively. $H_1$ describes the coupling between the two mechanical modes and the optomechanical interaction between particle~$A$ and the cavity,
\begin{align}
H_1 = g_{ab} a b^{\dagger} + g_{ba} a^{\dagger} b + g (a^{\dagger} c + a c^{\dagger}),
\end{align}
where the parameters $g_{ab} = g_1 - g_2$ and $g_{ba} = g_1 + g_2$ represent the asymmetric coupling strengths between the two particles~\cite{Metelmann2015,Fang2017,Reisenbauer2024,Liska2024}. The interaction becomes non-Hermitian when $\mathrm{Re}(g_2) \neq 0$, corresponding to unidirectional energy flow between the particles. The parameter $g$ denotes the linearized coupling rate between particle~$A$ and the cavity field.

In a frame rotating at the driving laser frequency $\omega_L$, the cavity frequency is replaced by the detuning parameter $\Delta = \omega_c - \omega_L$, and the free Hamiltonian becomes $H_0' = -\Delta\, c^{\dagger} c + \omega_a a^{\dagger} a + \omega_b b^{\dagger} b$. This transformation eliminates the fast optical oscillations. Under the rotating-wave approximation, rapidly oscillating counter-rotating terms such as $a c$ and $a^{\dagger} c^{\dagger}$ are neglected since their effects average out when the coupling strengths are small compared with the optical and mechanical frequencies. The effective Hamiltonian governing the system dynamics in the rotating frame is therefore~\cite{Liu2013,Clerk2010}
\begin{align}
H_{\mathrm{eff}} = & -\Delta\, c^{\dagger} c + \omega_a a^{\dagger} a + \omega_b b^{\dagger} b\notag\\
& + g_{ab} a b^{\dagger} + g_{ba} a^{\dagger} b + g (a^{\dagger} c + a c^{\dagger}).
\label{eq:H_eff}
\end{align}
The detuned cavity–mechanical coupling, the asymmetric inter-particle interaction, and the unidirectional phonon exchange characteristic of non-Hermitian dynamics are explicitly captured in Eq.~\eqref{eq:H_eff}.

Initially, both particle~$A$ and $B$ are in the thermal states with the average thermal phonon number $\overline{n}_a$ and $\overline{n}_b$. We achieve cooling of the target particle~$B$ through the cavity and the auxiliary particle~$A$. Considering the non-Hermitian interaction and the unavoidable intrinsic loss, the evolution of the whole system containing decay and damping can be described by the quantum master equation~\cite{GardinerZoller2004}
\begin{align}
\dot{\rho}_{\mathrm{sys}} = & \ \mathcal{L} \rho_{\mathrm{sys}}\notag\\
= &- i\, (H_{\mathrm{eff}} \rho_{\mathrm{sys}} - \rho_{\mathrm{sys}} H_{\mathrm{eff}}^{\dagger})\notag\\
& + i\, \mathrm{Tr}[\rho_{\mathrm{sys}} (H_{\mathrm{eff}} - H_{\mathrm{eff}}^{\dagger})] \rho_{\mathrm{sys}} + (\mathcal{L}_a + \mathcal{L}_b + \mathcal{L}_c) \rho_{\mathrm{sys}}, \notag\\
\mathcal{L}_a = & \ (1 + \overline{n}_a)\kappa_a D_{a}+\overline{n}_a \kappa_a D_{a^{\dagger}}, \notag\\
\mathcal{L}_b = & \ (1 + \overline{n}_b)\kappa_b D_{b}+\overline{n}_b \kappa_b D_{b^{\dagger}}, \notag\\
\mathcal{L}_c = & \ \gamma D_c.
\label{eq:three_mode_master}
\end{align}
where $D_x$ refers to the notation in the Lindblad form, which is expressed as $D_x(\rho) = 2x \rho x^{\dagger} - x^{\dagger} x \rho-\rho x^{\dagger} x$. The trace-correction term $i \,\mathrm{Tr}[\rho_{\mathrm{sys}} (H_{\mathrm{eff}} - H_{\mathrm{eff}}^{\dagger})] \rho_{\mathrm{sys}}$ is introduced to guaranty trace preservation during non-Hermitian evolution. $\kappa_a$ and $\kappa_b$ are the mechanical decay of particle~$A$ and $B$, respectively, and $\gamma$ is the intrinsic decay of the cavity. Eq.~\eqref{eq:three_mode_master} will be solved in both analytical and numerical methods, and the results will be discussed.

\section{\label{sec:level3}Analytical Solution}

In this section, we derive the analytical expressions governing the steady--state phonon numbers of the two mechanical modes. Starting from the complete three-mode master equation presented in Sec.~II, we first obtain an effective description of particle~$A$ by treating the cavity–$A$ subsystem as a linearized optomechanical pair. This allows us to identify an optically induced damping rate and an effective thermal occupation for mode~$a$. We then incorporate these effective parameters into a reduced two-mode non-Hermitian model for particles~$A$ and~$B$, from which closed dynamical equations for the phonon numbers can be obtained.  

Throughout this section, we make use of the fact that, under the bad-cavity limit and the harmonic oscillator approximation, mode~$a$ is cooled to a sufficiently small amplitude such that the nonlinear trace-correction term  $i\, \mathrm{Tr}[\rho_{\mathrm{sys}} (H_{\mathrm{eff}} - H_{\mathrm{eff}}^{\dagger})] \rho_{\mathrm{sys}}$ in the original master equation contributes negligibly to the dynamics (see Appendix~A for the formal justification).

We consider the coupled cavity–$A$ subsystem prior to introducing particle~$B$.  
In a frame rotating at the cavity drive frequency, the standard linearized optomechanical Hamiltonian reads~\cite{Liu2013,Clerk2010}
\begin{align}
H_{c-A}
= -\Delta\, c^{\dagger}c 
+ \omega_{a}\, a^{\dagger}a 
+ g (a^{\dagger}c + ac^{\dagger}).
\end{align}
The cavity induces anti-Stokes and Stokes scattering rates
\begin{align}
A_{-} &= g^{2}\frac{\gamma}{(\Delta-\omega_{a})^{2}+(\gamma/2)^{2}},\\
A_{+} &= g^{2}\frac{\gamma}{(\Delta+\omega_{a})^{2}+(\gamma/2)^{2}},
\end{align}
which respectively cool and heat the mechanical mode.  
The optically induced damping rate and effective occupation of the optical reservoir are therefore
\begin{align}
\Gamma_{\mathrm{opt}} &= A_{-}-A_{+},\\
n_{\mathrm{opt}} &= \frac{A_{+}}{A_{-}-A_{+}}.
\end{align}
Combining optical and intrinsic mechanical dissipation, the total effective damping rate and total effective bath occupation of particle~$A$ become
\begin{align}
\kappa_{a}^{\mathrm{(eff)}} &= \kappa_{a} + \Gamma_{\mathrm{opt}},\\
\overline{n}_{a}^{\mathrm{(eff)}} &= 
\frac{\kappa_{a}\overline{n}_{a} + \Gamma_{\mathrm{opt}} n_{\mathrm{opt}}}
{\kappa_{a}^{\mathrm{(eff)}}}.
\end{align}
These expressions provide an explicit mapping from the physical cavity parameters 
$(\gamma,\,\Delta,\,g)$
to an effective thermal environment for particle~$A$. For the convenience of the notation, we rename $\kappa_{a}^{\mathrm{(eff)}} \rightarrow \kappa_{a}'$ , $\overline{n}_{a}^{\mathrm{(eff)}} \rightarrow \overline{n}_{a}'$.

With the cavity eliminated in favor of an effective bath for mode $a$, we now consider the remaining two-mechanical-mode system composed of particles~$A$ and~$B$.  
The Hamiltonian is
\begin{align}
H_{A-B}
= \omega_{a} a^{\dagger}a + \omega_{b} b^{\dagger}b
+ g_{ab} a b^{\dagger}
+ g_{ba} a^{\dagger} b.
\end{align}
Mode $a$ now couples to the effective bath characterized by
$(\kappa_{a}',\, \overline{n}_{a}')$, 
while mode $b$ retains its intrinsic dissipation 
$(\kappa_{b},\, \overline{n}_{b})$.  
The master equation for the reduced density matrix $\rho_{a,b}$ is
\begin{align}
\dot{\rho}_{a,b}
=&\ \mathcal{L}'\rho_{a,b}\notag\\
=& -i\,(H_{A-B}\rho_{a,b} - \rho_{a,b}H_{A-B}^{\dagger}) + (\mathcal{L}_a' + \mathcal{L}_b)\rho_{a,b},
\label{eq:two_mode_master}
\end{align}
where $\mathcal{L}_a' =  \ (1 + \overline{n}_a')\kappa_a' D_{a}+\overline{n}_a' \kappa_a' D_{a^{\dagger}}$.

Applying the adjoint Liouvillian corresponding to Eq.~\eqref{eq:two_mode_master} to the number operators $n_{a}=a^{\dagger}a$ and $n_{b}=b^{\dagger}b$ yields coupled nonlinear differential equations~\cite{GardinerZoller2004} (see Appendix~B for the full derivation):
\begin{align}
\frac{d}{dt}n_{a}
=& \;\frac{2(g_{1}+g_{2})^{2}}{\kappa_{a}'+\kappa_{b}}
(1+n_{a})n_{b} - \frac{2(g_{1}-g_{2})^{2}}{\kappa_{a}'+\kappa_{b}}
n_{a}(1+n_{b}) \notag\\
&+ \kappa_{a}'\big(\overline{n}_{a}'-n_{a}\big), \notag\\
\frac{d}{dt}n_{b}
=& -\frac{2(g_{1}+g_{2})^{2}}{\kappa_{a}'+\kappa_{b}}
(1+n_{a})n_{b} + \frac{2(g_{1}-g_{2})^{2}}{\kappa_{a}'+\kappa_{b}}
n_{a}(1+n_{b}) \notag\\
&+ \kappa_{b}\big(\overline{n}_{b}-n_{b}\big).
\label{eq:adjoint}
\end{align}
At the steady state, i.e., when $d n_a/dt = d n_b/dt = 0$, taking the quantum expectation values of both sides and employing the mean-field approximation $\langle n_a n_b \rangle = \langle n_a \rangle \langle n_b \rangle$, we can solve the steady--state for the average phonon number of particle~$B$ from a quadratic equation written as
\begin{align}
A\, \langle n_{b}\rangle^{2} + B\, \langle n_{b}\rangle + C = 0,
\label{eq:n_b}
\end{align}
with coefficients
\begin{align}
A =& \;\frac{8g_1 g_2\kappa_b}{\kappa_{a}'(\kappa_{a}'+\kappa_b)}\notag\\
B =& \; - \frac{8g_1 g_2}{\kappa_{a}'+\kappa_b}\Bigl(\overline{n}_{a}' + \frac{\kappa_b}{\kappa_{a}'}\overline{n}_b\Bigr)\notag\\
    &- \frac{2(g_1+g_2)^2}{\kappa_{a}'+\kappa_b}
     - \frac{2(g_1-g_2)^2}{\kappa_{a}'+\kappa_b}\cdot\frac{\kappa_b}{\kappa_{a}'} - \kappa_b\notag\\
C =& \;\frac{2(g_1-g_2)^2}{\kappa_{a}'+\kappa_b}\,
       \Bigl(\overline{n}_{a}' + \frac{\kappa_b}{\kappa_{a}'}\overline{n}_b\Bigr)
     + \kappa_b \overline{n}_b .
\end{align}

The steady--state for the average phonon number of particle~$A$ is derived analogously and provided in Appendix~C. We numerically solved \eqref{eq:adjoint} for both particles, excluding nonphysical solutions, and analyzed the steady--state phonon populations $\langle n_a \rangle$ and $\langle n_b \rangle$ as functions of the nonreciprocal coupling parameter $K_g = (g_1 + g_2)/(g_1 - g_2)$ ($g_1 \neq g_2$). This parameter characterizes the degree of nonreciprocity between particle~$A$ and particle~$B$, with a larger $K_g$ representing stronger unidirection energy transfer from particle~$B$ to particle~$A$ within the same interaction timescale. 

\begin{figure}[htbp]
\includegraphics[width=0.5\textwidth]{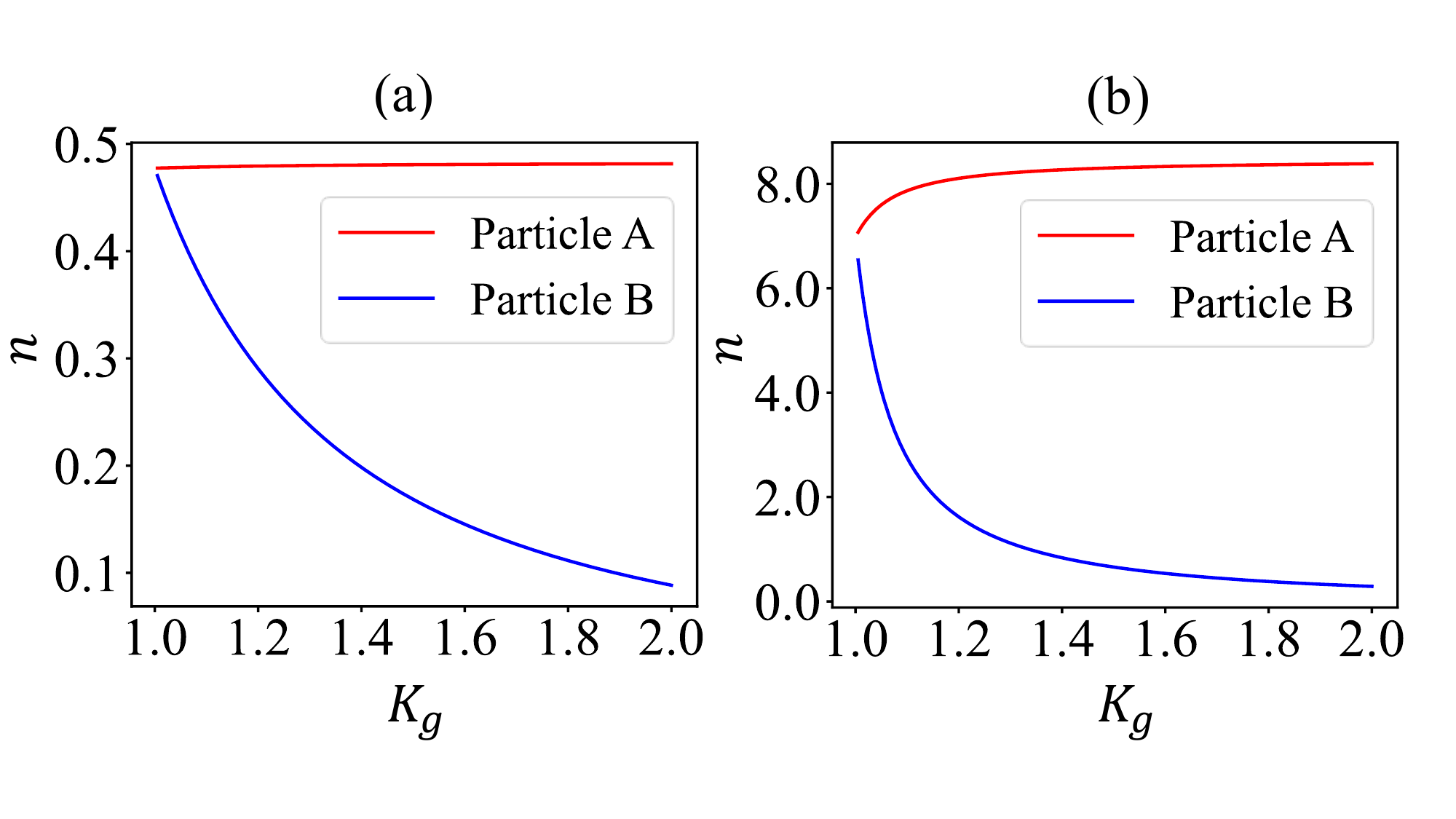}
\caption{\label{fig:fig2} Steady--state phonon numbers of particles $A$ and $B$ versus the nonreciprocal coupling parameter $K_g$. All parameters are expressed in dimensionless units with $\omega_a=\omega_b=\Delta=1$. The remaining parameters are fixed as $\gamma = 1$, $\overline{n}_a = \overline{n}_b = 20$, $\kappa_a = \kappa_b = 10^{-4}$, and $g_2 = 0.1$. The reciprocal coupling coefficient is chosen as $g_1 = g_2 (K_g + 1)/(K_g - 1)$ so that $K_g>1$ corresponds to an increasingly nonreciprocal interaction. Particle~$A$ and $B$ correspond to the red and blue lines, respectively. (a) Results for cavity-$A$ coupling $g=0.05$. Here $n_A$ remains almost constant, while $n_B$ decreases significantly with $K_g$ due to efficient directional energy transfer into the strongly cooled particle~$A$. (b) Results for weaker coupling $g=0.01$. In this regime $n_B$ continues to decrease exponentially with $K_g$, whereas $n_A$ increases slowly and reaches values one order of magnitude larger than in panel~(a), reflecting phonon accumulation at particle $A$. In both panels, $n_A$ and $n_B$ approach each other as $K_g\to1^{+}$, consistent with the reciprocal coupling limit.}
\end{figure}

In our calculations, the characteristic mechanical frequency used lies in the range of $10^{5}$--$10^{6}\,\mathrm{Hz}$, consistent with experimentally measured center-of-mass and torsional-mode frequencies of optically levitated nanoparticles in cavity and tweezer-based traps~\cite{Reisenbauer2024,Liska2024}. For convenience, all the parameters in the calculation are expressed in dimensionless units, where the mechanical frequencies are normalized as $\omega_a = \omega_b = \Delta = 1$. The cavity–$A$ coupling strengths $g = 0.01$--$0.05$ correspond to linearized optomechanical coupling rates of approximately $10^{3}$--$10^{4}\,\mathrm{Hz}$, and the mechanical damping rates $\kappa_{a} = \kappa_{b} = 10^{-4}$ represent intrinsic line widths of $10$--$100\,\mathrm{Hz}$ typical for levitated nanoparticles at high vacuum. The thermal occupations $\overline{n}_a = \overline{n}_b=20$ correspond to effective mode temperatures in the millikelvin regime following precooling through feedback or cavity-assisted schemes. Recent torsional-mode implementations have demonstrated that nonreciprocal mechanical coupling can reach relatively large values of $K_g$~\cite{Reisenbauer2024,Liska2024}, highlighting the feasibility of accessing strongly asymmetric interaction regimes in experiments.

 


As shown in Fig.~2, the analytical steady--state solutions exhibit a clear dependence of the phonon populations on the degree of nonreciprocity. When the cavity-$A$ coupling is relatively strong as in Fig.~2(a), particle~$A$ is efficiently cooled and therefore remains nearly insensitive to variations in $K_g$, whereas particle~$B$ experiences a pronounced reduction in its phonon number due to the increasingly directional energy flow from $B$ to $A$. In contrast, for weaker cavity coupling as in Fig.~2(b), the reduced optical damping of particle~$A$ results in gradual phonon accumulation as $K_g$ increases, while particle~$B$ still undergoes enhanced cooling under stronger nonreciprocity. In both regimes, the limit $K_g \to 1^{+}$ leads to $n_A \approx n_B$, consistent with the restoration of reciprocal coupling. These results collectively demonstrate that nonreciprocal interactions offer a robust mechanism for improving the remote cooling of particle~$B$, with the behavior of particle~$A$ primarily governed by the strength of its cavity--induced dissipation.

\section{\label{sec:level4}Numerical Simulation}

To verify the analytical results and investigate the dynamical cooling behavior of the coupled system, we numerically solve the full time-dependent master equation for the total density matrix $\rho_{\mathrm{sys}}(t)$ as described in Eq.~\eqref{eq:three_mode_master}. The time evolution of $\rho_{\mathrm{sys}}(t)$ is computed using the Runge--Kutta algorithm, which allows high-order integration accuracy and flexible adaptive time stepping. The Hilbert space of each mode is truncated to a finite dimension to ensure numerical convergence while keeping computational costs manageable~\cite{GardinerZoller2004,BreuerPetruccione2002,Johansson2012,Johansson2013}. The initial state is chosen as $|0\rangle_A |2\rangle_B |0\rangle_C$, corresponding to the case where only particle~$B$ initially possesses vibrational excitations, while particle~$A$ and the cavity mode are both in their ground states.

During the time evolution, we calculate the expectation values of the phonon number operators $n_A = \mathrm{Tr}[\rho_{\mathrm{sys}} a^{\dagger} a]$, $n_B = \mathrm{Tr}[\rho_{\mathrm{sys}} b^{\dagger} b]$, and $n_C = \mathrm{Tr}[\rho_{\mathrm{sys}} c^{\dagger} c]$ to monitor the cooling process of each mode~\cite{Manzano2020}. To guarantee the physical validity of the simulation, the trace of $\rho_{\mathrm{sys}}$, the minimum diagonal element, and the smallest eigenvalue of its Hermitian part are also evaluated throughout the evolution to verify trace preservation, positivity, and numerical stability of the density matrix~\cite{SchackBrun1997,Rivas2014}.

\begin{figure}[htbp]
\includegraphics[width=0.5\textwidth]{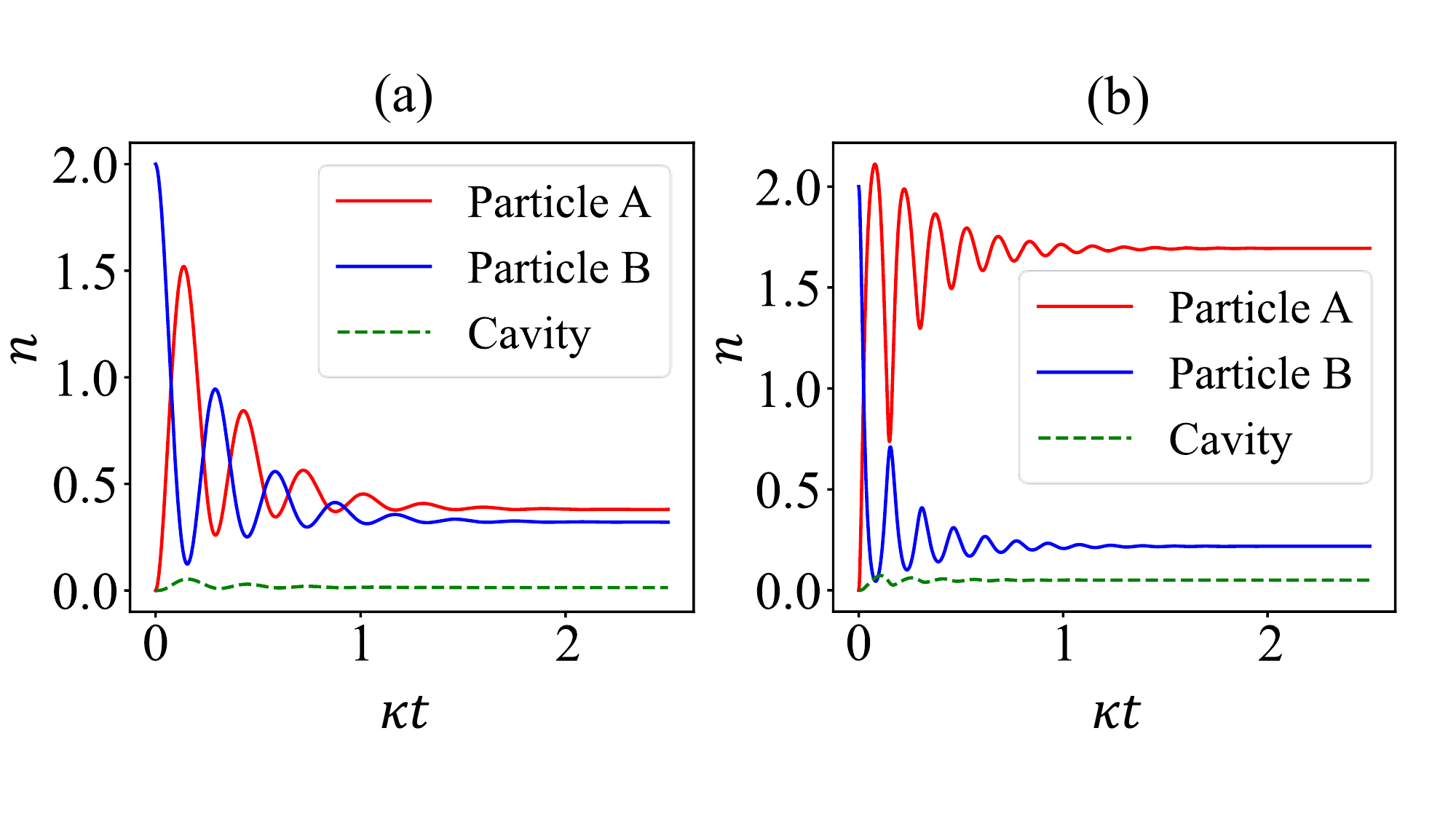}
\caption{\label{fig:fig3}
Time evolution of the phonon number $n$ for particles $A$ and $B$. particle~$B$ is the target to be cooled, while particle~$A$ acts as the auxiliary system mediating the cooling process. The initial phonon numbers of particles $A$ and $B$ are $0$ and $2$, respectively. The parameters are $g = 0.1$, $g_2 = 0.1$, $\gamma = 1.0$, $\kappa = \kappa_a = \kappa_b = 0.01$, and $\overline{n}_a = \overline{n}_b = 1$. Particle~$A$, $B$, and the cavity correspond to the red solid, blue solid, and green dashed lines, respectively. (a) $g_1 = 0.11$ ($K_g = 1.1$). The steady--state phonon numbers are $n_B = 0.32$ and $n_A = 0.38$. (b) $g_1 = 0.40$ ($K_g = 1.4$). The steady--state phonon numbers are $n_B = 0.22$ and $n_A = 1.69$. As $K_g$ increases, the nonreciprocity strengthens, leading to lower $n_B$ and higher $n_A$, 
which highlights the directional energy transfer and enhanced non-Hermitian cooling efficiency.}
\end{figure}

As shown in Fig.~3, the time evolution of the phonon numbers for particles $A$ and $B$ under two distinct coupling conditions is plotted. In both cases, the system evolves from the initial nonequilibrium configuration toward a steady state, where the phonon populations of the two particles exhibit a clear asymmetry. The target particle~$B$ continuously loses energy to particle~$A$ through the nonreciprocal interaction channel, leading to a reduction in its steady--state phonon number. By contrast, particle~$A$ accumulates energy during the same process, resulting in an increased phonon occupation.

The comparison between Fig.~3(a) and Fig.~3(b) reveals that increasing the nonreciprocal parameter $K_g$ 
not only enhances the unidirectional energy transfer but also substantially reduces the steady--state phonon population of the target particle~$B$. Although relatively small parameters are chosen here due to the numerical cost of directly integrating the full master equation after Hilbert-space truncation, the observed cooling contrast between different $K_g$ values is already significant. Under realistic experimental parameters with stronger coupling and higher cavity quality factors, the cooling asymmetry and overall reduction in $\langle n_B \rangle$ are expected to be even more pronounced. These results confirm that non-Hermitian interactions can effectively facilitate asymmetric energy dissipation and enable sympathetic cooling below the conventional optomechanical limits.

For larger parameter regimes where the Hilbert-space dimension becomes significantly larger, direct time-domain integration of the master equation is computationally demanding and memory-intensive. In this case, the steady--state phonon number is obtained by solving the stationary master equation $\mathcal{L} \rho_{\mathrm{ss}} = 0$ using sparse-matrix techniques~\cite{BreuerPetruccione2002,Johansson2013}. Specifically, we construct the full Liouvillian superoperator $\mathcal{L}$ in the column-stacking representation~\cite{Prosen2010,White2004}, where both coherent and dissipative contributions are included in a sparse compressed sparse row format. The steady--state density matrix $\rho_{\mathrm{ss}}$ is then determined by replacing one row of $\mathcal{L}$ with the trace constraint $\mathrm{Tr}(\rho_{\mathrm{ss}}) = 1$~\cite{BarnettRadmore1997} and solving the resulting linear system. After the numerical solution is obtained, the density matrix is Hermitized and normalized to guarantee physical validity (trace preservation and positivity). The steady--state phonon numbers $\langle n_A \rangle$ and $\langle n_B \rangle$ are then computed from $\langle n_j \rangle = \mathrm{Tr}[\rho_{\mathrm{ss}} a_j^{\dagger} a_j]$ ($j = A, B$).

\begin{figure}[htbp]
\includegraphics[width=0.5\textwidth]{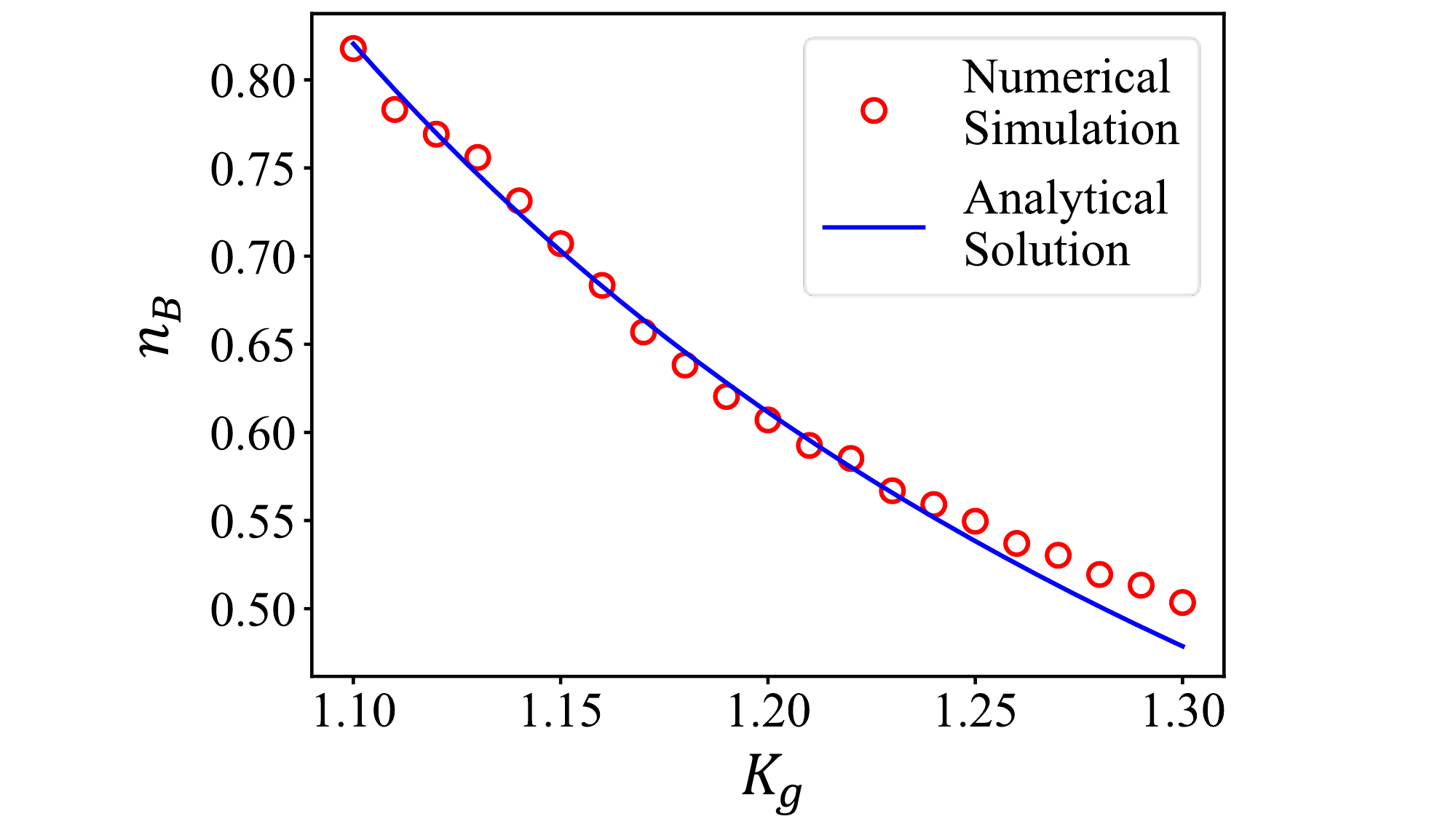}
\caption{\label{fig:fig4}
Comparison between the analytical and numerical steady--state phonon numbers of particle~$B$ as a function of the nonreciprocal coupling parameter $K_g$. The red dots represent the results obtained from the sparse-Liouvillian steady--state solver, and the blue solid line shows the analytical solution based on Eq.~\eqref{eq:adjoint}. $g_2 = 0.1$, $g_1 = g_2(K_g + 1)/(K_g - 1)$, $g = 0.05$, $\kappa_a = \kappa_b = 10^{-4}$, $\gamma = 2$, and $\overline{n}_a = \overline{n}_b = 20$.}
\end{figure}

We have compared the analytical steady--state phonon occupations with the numerical results obtained from the full three--mode Liouvillian simulation, as shown in Fig.~4. In the weakly nonreciprocal regime, where $K_g$ is only slightly above unity, the agreement between the two approaches is very good, confirming the consistency of the effective--bath reduction and the reduced two--mode non-Hermitian model. As $K_g$ increases, the analytical solution predicts enhanced directional energy transfer from particle $B$ to particle $A$, and while the numerical results follow the same qualitative behavior, quantitative deviations emerge once the phonon population of particle $A$ grows sufficiently large. In this regime the truncated Hilbert space cannot fully capture the high--excitation tails, consistent with convergence diagnostics, although the numerically stable region still exhibits clear agreement with the theoretical trend. Overall, the comparison demonstrates that the analytical model faithfully describes the key features of non-Hermitian cooling and its dependence on the nonreciprocity parameter $K_g$.

\section{\label{sec:level5}Conclusion}

We have theoretically proposed and analyzed a sympathetic cooling scheme through nonreciprocal coupling between two optically levitated nanoparticles. In this system, only one particle is directly coupled to an optical cavity, while the other is cooled indirectly through a nonreciprocal mechanical interaction. By introducing asymmetric coupling coefficients, we derived an effective non-Hermitian Hamiltonian and obtained analytical expressions for the steady--state phonon numbers. Both analytical and numerical results consistently demonstrate that increasing the nonreciprocity parameter $K_g$ enhances directional energy transfer from the target particle to the auxiliary particle, resulting in a lower steady--state phonon number of the target particle and more efficient cooling.

The key insight gained from this study is that by introducing an auxiliary particle that interacts nonreciprocally with the target particle, one can engineer a synthetic energy flow channel that enables cooling the target particle mode to a lower effective temperature. This mechanism fundamentally differs from traditional cavity-assisted sideband cooling, as it does not solely rely on cavity dissipation but exploits the asymmetry of coupling to break detailed balance and enable directional heat transfer.

Our results provide a new perspective for controlling mechanical motion in levitated optomechanical systems through tunable nonreciprocal interactions. While demonstrated for a two-particle configuration, the scheme naturally extends to multi-particle networks in which synthetic gauge fields~\cite{Fang2017,Metelmann2015}, engineered dissipation~\cite{Metelmann2014}, or phase-controlled optical feedback~\cite{Kippenberg2005,Ruesink2016} can establish spatially distributed nonreciprocal couplings. In particular, arranging nanoparticles into a unidirectional non-Hermitian chain enables cascaded or even exponentially enhanced cooling~\cite{Xu2024,Clerk2020}, as suggested by recent theoretical studies. Such chain geometries amplify directional energy flow and can substantially deepen the achievable cooling of mechanical modes beyond what is possible with reciprocal or purely local interactions.

The essential ingredients for implementing these schemes—optical phase control, spatial-mode engineering, and feedback-assisted interaction tuning—are already available in contemporary levitated optomechanics experiments. Optical tweezers arrays and cavity-assisted traps provide fine control over trapping geometries and phases~\cite{Frangeskou2018,Barker2010}, while demonstrated nonreciprocal torsional couplings indicate the feasibility of accessing strongly asymmetric regimes~\cite{Reisenbauer2024, Liska2024}. These experimental capabilities suggest that the proposed mechanism can be realized with realistic parameters and may enable robust, directional cooling channels without the need for extremely high cavity finesse. Overall, this work advances the theoretical foundation for exploiting non-Hermitian effects in levitated systems and highlights their potential for scalable quantum control, ultra-low-noise sensing, and many-body quantum physics.

\begin{acknowledgments}
This work is supported by the National Natural Science Foundation of China (Grant No. 12441502) and  the Beijing Institute of Technology Research Fund Program under Grant No. 2024CX01015.
\end{acknowledgments}

\nocite{*}

\appendix
\section{Dispose of the Trace-correction Term}

Here, we discuss the trace-correction term $i \,\mathrm{Tr}[\rho_{\mathrm{sys}} (H_{\mathrm{eff}} - H_{\mathrm{eff}}^{\dagger})] \rho_{\mathrm{sys}}$ in Eq.~\eqref{eq:three_mode_master} in the main text. Substituting $H_{\mathrm{eff}} - H_{\mathrm{eff}}^{\dagger} = 2 g_2 (a^{\dagger} b - a b^{\dagger})$, we obtain
\begin{align}
i \,\mathrm{Tr}[\rho_{\mathrm{sys}} (H_{\mathrm{eff}} - H_{\mathrm{eff}}^{\dagger})] \rho_{\mathrm{sys}} = 2 i g_2 \mathrm{Tr}[\rho (a^{\dagger} b - a b^{\dagger})] \rho .
\label{eq:A1}
\end{align}

Since particle~$A$ is coupled to the optical cavity, under the bad-cavity limit (which is satisfied by the parameters we adopt), the mode of particle~$A$ is strongly cooled and can be regarded as a small quantity $\alpha$, which can be written as $\alpha = |\alpha| e^{i \theta}$. With this approximation, Eq.~\eqref{eq:A1} can be rewritten as
\begin{align}
i \,\mathrm{Tr}[\rho_{\mathrm{sys}} (H_{\mathrm{eff}} - H_{\mathrm{eff}}^{\dagger})] \rho_{\mathrm{sys}} & = 2 i g_2 \mathrm{Tr}[\rho (\alpha^* b - \alpha b^{\dagger})] \rho\notag\\
& = 2 i g_2 |\alpha| \mathrm{Tr}[\rho (e^{-i \theta} b - e^{i \theta} b^{\dagger})] \rho .
\label{eq:A2}
\end{align}

By introducing a phase transformation $b' = b e^{-i \theta}$, so that $b'^{\dagger} = b^{\dagger} e^{i \theta}$, Eq.~\eqref{eq:A2} can be further simplified as
\begin{align}
i \,\mathrm{Tr}[\rho_{\mathrm{sys}} (H_{\mathrm{eff}} - H_{\mathrm{eff}}^{\dagger})] \rho_{\mathrm{sys}} = 2 i g_2 |\alpha| \mathrm{Tr}[\rho (b' - b'^{\dagger})] \rho .
\end{align}
Noting that $b$ and $b^{\dagger}$ are the annihilation and creation operators of particle~$B$, we have $b - b^{\dagger} = i k p_b$, where $k$ is a constant and $p_b$ is the momentum operator of particle~$B$. Under the harmonic oscillator approximation, the expectation value of $p_b$ is zero. Because $b'$ differs from $b$ only by a phase factor $e^{-i \theta}$, we obtain
\begin{align}
i \,\mathrm{Tr}[\rho_{\mathrm{sys}} (H_{\mathrm{eff}} - H_{\mathrm{eff}}^{\dagger})] \rho_{\mathrm{sys}} = 2 i g_2 |\alpha| \mathrm{Tr}[\rho (b' - b'^{\dagger})] \rho = 0 .
\end{align}

Therefore, we have demonstrated that, under the harmonic oscillator approximation and the bad-cavity limit, $i \,\mathrm{Tr}[\rho_{\mathrm{sys}} (H_{\mathrm{eff}} - H_{\mathrm{eff}}^{\dagger})] \rho_{\mathrm{sys}}$ vanishes.

\section{Derivation of the Adjoint Equations for the Number Operators}

In this section, we derive the adjoint equations for the number operators from the master equation for $\rho_{a,b}$. First, we divide $H_{A-B}$ into $H_{\mathrm{sys}} = \omega_{a} a^{\dagger}a + \omega_{b} b^{\dagger}b$ and $H_{\mathrm{int}} = g_{ab} a b^{\dagger} + g_{ba} a^{\dagger} b$. Then the total Liouvillian $\mathcal{L}'$ in Eq.~\eqref{eq:two_mode_master} can be decomposed as $\mathcal{L}' = \mathcal{L}_{\mathrm{sys}} + \mathcal{L}_{\mathrm{int}}$, where
\begin{align}
\mathcal{L}_{\mathrm{sys}}\rho & = -i [H_{\mathrm{sys}}, \rho] + \mathcal{L}_a'\rho + \mathcal{L}_b\rho, \notag\\
\mathcal{L}_{\mathrm{int}}\rho & = -i (H_{\mathrm{int}} \rho - \rho H_{\mathrm{int}}^{\dagger}).
\end{align}

Performing a second-order perturbative expansion of the master equation Eq.~\eqref{eq:two_mode_master}, we get
\begin{align}
\dot{\rho}_{a,b}(t) = & \ \mathcal{L}_{\mathrm{sys}} \rho_{a,b}(t) + \mathcal{L}_{\mathrm{int}} \int_0^{\infty} dt' e^{\mathcal{L}_{\mathrm{sys}}t'} \mathcal{L}_{\mathrm{int}}\, \rho_{a,b}(t-t').
\label{eq:B2}
\end{align}
Substituting the expression of $\mathcal{L}_{\mathrm{int}}$ into Eq.~\eqref{eq:B2} and expanding step by step, we first obtain
\begin{align}
&\mathcal{L}_{\mathrm{int}} \int_0^{\infty} dt' e^{\mathcal{L}_{\mathrm{sys}}t'} \mathcal{L}_{\mathrm{int}} \rho_{a,b}(t-t')=\notag\\
&-i H_{\mathrm{int}} \int_0^{\infty} dt' e^{\mathcal{L}_{\mathrm{sys}}t'} \mathcal{L}_{\mathrm{int}} \rho_{a,b}(t-t')\notag\\
&+i \int_0^{\infty} dt' e^{\mathcal{L}_{\mathrm{sys}}t'} \mathcal{L}_{\mathrm{int}} \rho_{a,b}(t-t') H_{\mathrm{int}}^{\dagger}.
\end{align}
Meanwhile,
\begin{align}
\mathcal{L}_{\mathrm{int}} \rho_{a,b}(t-t') =& - i H_{\mathrm{int}} \rho_{a,b}(t-t') + i \rho_{a,b}(t-t') H_{\mathrm{int}}^{\dagger}.
\end{align}

By applying the transformation between the Schrödinger and Heisenberg pictures and using the Markov approximation, we have
\begin{align}
e^{\mathcal{L}_{\mathrm{sys}}t'} \mathcal{L}_{\mathrm{int}} \rho_{a,b}(t-t')=&-ie^{\mathcal{L}_{\mathrm{sys}}^{\dagger}t'}(H_{\mathrm{int}}) \rho_{a,b}(t)\notag\\
&+i \rho_{a,b}(t) e^{\mathcal{L}_{\mathrm{sys}}^{\dagger}t'}(H_{\mathrm{int}}^\dagger).
\end{align}
Thus, the reduced master equation can be written as
\begin{align}
\dot{\rho}_{a,b}(t)=&-i [H_{\mathrm{sys}}, \rho_{a,b}(t)]\notag\\
&- \int_0^{\infty} dt' H_{\mathrm{int}} e^{\mathcal{L}_{\mathrm{sys}}^{\dagger}t'}(H_{\mathrm{int}}) \rho_{a,b}(t)\notag\\
&+ \int_0^{\infty} dt' H_{\mathrm{int}} \rho_{a,b}(t) e^{\mathcal{L}_{\mathrm{sys}}^{\dagger}t'}(H_{\mathrm{int}}^{\dagger})\notag\\
&+ \int_0^{\infty} dt' e^{\mathcal{L}_{\mathrm{sys}}^{\dagger}t'}(H_{\mathrm{int}}) \rho_{a,b}(t) H_{\mathrm{int}}^{\dagger}\notag\\
&- \int_0^{\infty} dt' \rho_{a,b}(t) e^{\mathcal{L}_{\mathrm{sys}}^{\dagger}t'}(H_{\mathrm{int}}^{\dagger}) H_{\mathrm{int}}^{\dagger}.
\label{eq:B6}
\end{align}

First, we evaluate
\begin{align}
e^{\mathcal{L}_{\mathrm{sys}}^{\dagger}t'}(H_{\mathrm{int}})=&\ (g_1 + g_2) (e^{\mathcal{L}_{\mathrm{sys}}^{\dagger}t'} a^\dagger)(e^{\mathcal{L}_{\mathrm{sys}}^{\dagger}t'} b)\notag\\
&+(g_1 - g_2) (e^{\mathcal{L}_{\mathrm{sys}}^{\dagger}t'} a)(e^{\mathcal{L}_{\mathrm{sys}}^{\dagger}t'} b^{\dagger}).
\label{eq:B7}
\end{align}
The action of $\mathcal{L}_{\mathrm{sys}}^{\dagger}$ on $a$ is given by
\begin{align}
\mathcal{L}_{\mathrm{sys}}^{\dagger}a=&\ i[H_{\mathrm{sys}},a]+\mathcal{L}_a'^{\dagger}a+\mathcal{L}_b^{\dagger}a\notag\\
=&\ i[\omega_a a^\dagger a,a]+(1+\overline{n}_a')\kappa_a' (2 a^{\dagger} a a-a^{\dagger} a a-a a^{\dagger} a)\notag\\
&+\overline{n}_a' \kappa_a' (2 a a a^{\dagger}-a a^{\dagger} a-a a a^{\dagger})\notag\\
=&-(i \omega_a +\kappa_a') a.
\end{align}
Hence, $e^{\mathcal{L}_{\mathrm{sys}}^{\dagger}t'} a=e^{-(\kappa_a'+i \omega_a)t'} a$. Similarly, we get $e^{\mathcal{L}_{\mathrm{sys}}^{\dagger}t'} a^\dagger=e^{-(\kappa_a'-i \omega_a)t'} a^\dagger$, $e^{\mathcal{L}_{\mathrm{sys}}^{\dagger}t'} b=e^{-(\kappa_b+i \omega_b)t'} b$, $e^{\mathcal{L}_{\mathrm{sys}}^{\dagger}t'} b^\dagger=e^{-(\kappa_b-i \omega_b)t'} b^\dagger$.
Substituting these relations into Eq.~\eqref{eq:B7}, we obtain
\begin{align}
e^{\mathcal{L}_{\mathrm{sys}}^{\dagger}t'}(H_{\mathrm{int}})=&(g_1 + g_2) e^{- \big( (\kappa_a'+\kappa_b)-i(\omega_a-\omega_b) \big) t'} a^\dagger b\notag\\
&+(g_1 - g_2) e^{- \big( (\kappa_a'+\kappa_b)+i(\omega_a-\omega_b) \big) t'} a b^\dagger,
\label{eq:B9}
\end{align}
where fast-oscillating terms such as $a^{2}$ and $a^{\dagger 2}$ have been neglected.
Similarly,
\begin{align}
e^{\mathcal{L}_{\mathrm{sys}}^{\dagger}t'}(H_{\mathrm{int}}^\dagger)=&(g_1 + g_2) e^{- \big( (\kappa_a'+\kappa_b)+i(\omega_a-\omega_b) \big) t'} a b^{\dagger}\notag\\
&+(g_1 - g_2) e^{- \big( (\kappa_a'+\kappa_b)-i(\omega_a-\omega_b) \big) t'} a^{\dagger} b.
\label{eq:B10}
\end{align}  

In the following analysis, the system frequencies are set as $\omega_a = \omega_b$. Using Eqs.~\eqref{eq:B9} and \eqref{eq:B10}, each term in Eq.~\eqref{eq:B6} can be rewritten as
\begin{align}
&- \int_0^{\infty} dt' H_{\mathrm{int}} e^{\mathcal{L}_{\mathrm{sys}}^{\dagger}t'}(H_{\mathrm{int}}) \rho_{a,b}(t)=\notag\\
&-\frac{g_1^2 - g_2^2}{\kappa_a'+\kappa_b} (a a^\dagger b^\dagger b + a^\dagger a b b^\dagger) \rho_{a,b}(t),
\label{eq:B11}\\
& \int_0^{\infty} dt' H_{\mathrm{int}} \rho_{a,b}(t) e^{\mathcal{L}_{\mathrm{sys}}^{\dagger}t'}(H_{\mathrm{int}}^{\dagger})=\notag\\
&\frac{1}{\kappa_a'+\kappa_b} \Big( (g_1 - g_2)^2 a b^\dagger \rho_{a,b}(t) a^\dagger b + (g_1 + g_2)^2 a^\dagger b \rho_{a,b}(t) a b^\dagger \Big),\\
& \int_0^{\infty} dt' e^{\mathcal{L}_{\mathrm{sys}}^{\dagger}t'}(H_{\mathrm{int}}) \rho_{a,b}(t) H_{\mathrm{int}}^{\dagger}\notag\\
&\frac{1}{\kappa_a'+\kappa_b} \Big( (g_1 + g_2)^2 a^\dagger b \rho_{a,b}(t) a b^\dagger + (g_1 - g_2)^2 a b^\dagger \rho_{a,b}(t) a^\dagger b \Big),\\
&- \int_0^{\infty} dt' \rho_{a,b}(t) e^{\mathcal{L}_{\mathrm{sys}}^{\dagger}t'}(H_{\mathrm{int}}^{\dagger}) H_{\mathrm{int}}^{\dagger}=\notag\\
&-\frac{g_1^2 - g_2^2}{\kappa_a'+\kappa_b} \rho_{a,b}(t) (a^\dagger a b b^\dagger + a a^\dagger b^\dagger b).
\label{eq:B14}
\end{align}

By substituting Eqs.~\eqref{eq:B11}--\eqref{eq:B14} into Eq.~\eqref{eq:B6}, we obtain the reduced master equation
\begin{align}
\dot{\rho}_{a,b}(t) = & -i \Big[ H_{\mathrm{sys}}' + \frac{4 g_2^2}{\kappa_a' + \kappa_b} a^{\dagger} a b^{\dagger} b,\, \rho_{a,b}(t) \Big] \notag\\
& + \mathcal{L}_a' \rho_{a,b}(t) + \mathcal{L}_b \rho_{a,b}(t) 
+ \mathcal{L}_{ab} \rho_{a,b}(t),
\label{eq:B15}
\end{align}
where the modified Hamiltonian $H_{\mathrm{sys}}'$ is given by
\begin{align}
H_{\mathrm{sys}}' = & \ \Big( \omega_a + \frac{2g_2^2 - 2g_1 g_2}{\kappa_a' + \kappa_b} \Big) a^{\dagger} a + \Big( \omega_b + \frac{2g_2^2 + 2g_1 g_2}{\kappa_a' + \kappa_b} \Big) b^{\dagger} b,
\end{align}
and the additional Liouvillian terms are defined as
\begin{align}
\mathcal{L}_{ab} &= \frac{(g_1 + g_2)^2}{\kappa'_a + \kappa_b} D_{a^{\dagger} b} 
+ \frac{(g_1 - g_2)^2}{\kappa'_a+ \kappa_b} D_{a b^{\dagger}}.
\end{align}
The general form of the adjoint equations, according to Eq.~\eqref{eq:B15}, is
\begin{align}
\frac{d}{dt} n_a =& (\mathcal{L}_{ab}^\dagger + \mathcal{L}_a'^\dagger) n_a\notag\\
=& \Big( \frac{(g_1 + g_2)^2}{\kappa_a'+\kappa_b} D_{a^\dagger b}^\dagger + \frac{(g_1 - g_2)^2}{\kappa_a'+\kappa_b} D_{a b^\dagger}^\dagger \Big) n_a \notag\\
&+ \Big( (1 + \overline{n}_a') \kappa_a' D_a^\dagger + \overline{n}_a' \kappa_a' D_{a^\dagger}^\dagger \Big) n_a, \notag\\
\frac{d}{dt} n_b =& (\mathcal{L}_{ab}^\dagger + \mathcal{L}_b^\dagger) n_b\notag\\
=& \Big( \frac{(g_1 + g_2)^2}{\kappa'_a+\kappa_b} D_{a^\dagger b}^\dagger + \frac{(g_1 - g_2)^2}{\kappa'_a+\kappa_b} D_{a b^\dagger}^\dagger \Big) n_b \notag\\
&+ \Big( ( 1 + \overline{n}_b) \kappa_b D_b^\dagger + \overline{n}_b \kappa_b D_{b^\dagger}^\dagger \Big) n_b.
\label{eq:C1}
\end{align}
Using the explicit forms of $D_x$ and $D_x^{\dagger}$, we obtain
\begin{align}
&D_{a^\dagger b}^\dagger(n_a) = 2(1 + n_a) n_b\ ,\ D_{a b^\dagger}^\dagger(n_a) = -2n_a (1 + n_b),\notag\\
&D_a^\dagger(n_a) = -2n_a\ ,\ D_{a^\dagger}^\dagger(n_a) = 2(1 + n_a),\notag\\
&D_{a^\dagger b}^\dagger(n_b) = -2(1 + n_a) n_b\ ,\ D_{a b^\dagger}^\dagger(n_b) = 2n_a (1 + n_b),\notag\\
&D_b^\dagger(n_b) = -2n_b\ ,\ D_{b^\dagger}^\dagger(n_b) = 2(1 + n_b).
\label{eq:C2}
\end{align}

Substituting Eq.~\eqref{eq:C2} into Eq.~\eqref{eq:C1}, the adjoint equations can be simplified to the form given in Eq.~\eqref{eq:adjoint} in the main text.

\section{Stationary State for the Average Phonon Number of Particle A}

Similar to Eq.~\eqref{eq:n_b} in the main text, the stationary state for the average phonon number of particle $A$ can be solved from a quadratic equation written as
\begin{align}
D {\langle n_a \rangle}^2 + E \langle n_a \rangle + F = 0
\end{align}
where
\begin{align}
D = & \;-\frac{8\,g_1 g_2\kappa_{a}'}{\kappa_b\,(\kappa_{a}'+\kappa_b)}\notag\\
E = & \;\frac{8 g_1 g_2}{\kappa_{a}'+\kappa_b}\Bigl(\frac{\kappa_{a}'}{\kappa_b}\overline{n}_{a}' + \overline{n}_b\Bigr)\notag\\
& - \frac{2(g_1+g_2)^2}{\kappa_{a}'+\kappa_b}\cdot\frac{\kappa_{a}'}{\kappa_b}-\frac{2(g_1-g_2)^2}{\kappa_{a}'+\kappa_b} - \kappa_{a}'\notag\\
F = & \;\frac{2(g_1+g_2)^2}{\kappa_{a}'+\kappa_b}\Bigl(\frac{\kappa_{a}'}{\kappa_b}\overline{n}_{a}' + \overline{n}_b\Bigr) + \kappa_{a}'\overline{n}_{a}' .
\end{align}

\bibliography{refs}

\end{document}